\documentclass[aps,prx,reprint,floatfix]{revtex4-1}

\usepackage{amsmath}    % need for subequations
\usepackage{amssymb}
\usepackage{graphicx}   % need for figures
\usepackage{color}      % use if color is used in text
\usepackage{hyperref}   % use for hypertext links, including those to external documents and URLs
\usepackage{dcolumn}

\def\lsco{La$_{2-x}$Sr$_x$CuO$_4$}
\def\lbco{La$_{2-x}$Ba$_x$CuO$_4$}

\def\ybco{YBa$_2$Cu$_3$O$_{6+x}$}

\def\bscco{Bi$_2$Sr$_2$CaCu$_2$O$_{8+\delta}$}

\begin{document}

\newcommand{\LSNO}{La$_{2-x}$Sr$_{x}$NiO$_{4}$}
\newcommand{\LSNOn}{La$_{1.75}$Sr$_{0.25}$NiO$_{4}$}

\title{Empirical case for two pseudogaps in cuprate superconductors}

\author{J. M. Tranquada}
\email{jtran@bnl.gov}
\affiliation{Condensed Matter Physics and Materials Science Division, Brookhaven National Laboratory, Upton, New York 11973, USA}

\date{\today} 

\begin{abstract}
Superconductivity in cuprates is achieved by doping holes into a correlated charge-transfer insulator.  While the correlated character of the parent insulator is now understood, there is no accepted theory for the ``normal'' state of the doped insulator.  I present a mostly empirical analysis of a large range of experimental characterizations, making the case for two pseudogaps: (1) a large pseudogap resulting from the competition between the energy of superexchange-coupled local Cu moments and the kinetic energy of doped holes; (2) a small pseudogap that results from dopant disorder and consequent variations in local charge density, leading to a distribution of local superconducting onset temperatures.  The large pseudogap closes as hole kinetic energy dominates at higher doping and the dynamic antiferromagnetic correlations become overdamped.  Establishing spatially-homogeneous $d$-wave superconductivity is limited by those regions with the weakest superconducting phase coherence, which tends to be limited by low-energy spin fluctuations.   The magnitude of the small pseudogap is correlated with the doping-dependent energy $E_{\rm cross}$ associated with the neck of the hour-glass dispersion of spin excitations.  The consequences of this picture are discussed.
\end{abstract}

\maketitle

\section{Introduction}

It is now well established that the parent compounds of cuprate superconductors are charge-transfer Mott-Hubbard insulators \cite{mott49,hubb64,zaan85}, where a consequence of the strong Coulomb repulsion in the half-filled Cu $3d_{x^2-y^2}$ orbital is antiferromagnetism driven by the superexchange mechanism \cite{ande59}.  Superconductivity with a high transition temperature, $T_c$, can be achieved by doping a sufficient density of holes into such a system.  We know that the resulting superconducting state involves pairing of holes \cite{goug87} with a $d$-wave symmetry \cite{tsue00}.  At the same time, there is no accepted understanding of the ``normal'' state from which the superconductivity develops, and finding an effective description remains a major challenge in the field \cite{keim15}.

Analysis often begins at finite doping.  Starting from a weak-coupling perspective, the depression of the measured density of charge carriers compared to the expectation for a model with no Coulomb repulsion has been characterized in terms of a pseudogap \cite{frie89}.  As the carrier density appears to decrease with cooling, the temperature-doping phase diagram is typically drawn with an effective phase boundary between high-temperature and pseudogap phases.  Despite the fact that the high-temperature phase is incoherent \cite{lee05}, it has been popular to propose mean-field pictures in which the onset of an order parameter that competes with superconductivity causes a reduction of the density of states at the Fermi energy \cite{slat51}.  Based on a weak-coupling interpretation of magnetic susceptibility, heat capacity, and resistivity measurements as a function of dopant-induced hole density $p$, Tallon and Loram \cite{tall01} proposed that the pseudogap closes at $p_c\sim0.2$.  A number of theorists have proposed that $p_c$ corresponds to a putative quantum critical point (QCP) associated with various possible order parameters \cite{varm97,cast96,sach99,chak01,sach03b,aban03}.  Recent experimental studies of quantities probing the electronic density of states have been interpreted as providing evidence for such a QCP \cite{bado16,mich19}.

From an alternative perspective, the pseudogap phenomena are a consequence of competing energies, rather than orders.  Following Anderson's lead \cite{ande07s}, one can start with the big energies in the problem and consider their impact.  There is now a great library of experimental results on hole-doped cuprates  \cite{timu99,norm05,hufn08,lee06}.  A recent review has made clear some of the apparent conflicts in interpreting the results from different experimental techniques \cite{keim15}.  Here, I attempt to combine the results and insights of many researchers to form an interpretation based on experimental results, with minimal reliance on theory, that overcomes many of the conflicts.  Rather than a review, this will be something of a meta-analysis.  My motivation has come from facing down inconsistencies of interpretation based on my own experiments \cite{li18}.

I will present the case that we need two pseudogaps \cite{taka01} in order to reconcile a variety of observations.  The large pseudogap results from a competition between superexchange-coupled Cu moments and the doped holes that would like to lower their kinetic energy by delocalizing.  Here the instantaneous correlations of neighboring spins are the key factor; these cannot be described by a static mean field theory. With doping, one observes a gradual reduction in the characteristic spin-excitation energy together with an evolution of the energy width associated with charge excitations.  This pseudogap tends to zero when the holes reach sufficient density to overdamp the remaining clusters of antiferromagnetic (AF) spins.  

The small pseudogap involves spatial variation in the local hole density, local development of superconducting coherence, and inhibition of long-range phase order due to regions with low-energy AF fluctuations.  The spatial disorder has long been clear from local probes such as scanning tunneling microscopy (STM) \cite{cren00,howa01,pan01} and nuclear magnetic resonance (NMR) studies \cite{sing02a,sing05}, and an appreciation for its impact has been growing \cite{pelc19}.  The maximal energy and temperature scales for local coherence and superconductivity fluctuations occur at $p\sim0.12$.  They are correlated with the doping dependence of the spin-excitation spectrum, which develops an ``hourglass'' form with a characteristic energy $E_{\rm cross}$ that appears to act as an upper limit for the pairing scale.  

I begin by considering the measured excitations of the lightly-doped Mott insulator, followed by a simple-minded theoretical analysis in Sec.~III.  The large pseudogap and its doping dependence are described in Sec.~IV.  The analysis of the small pseudogap is presented in Sec.~V.  Finally, the implications of this analysis are discussed in Sec.~V, with a brief summary in Sec.~VI.

\section{Doping a Mott insulator}

Three decades of experiments have taught us quite a bit about cuprates.  The undoped parent compounds, such as La$_2$CuO$_4$, are antiferromagnetic (AF) insulators \cite{kast98} with a superexchange coupling in the range of 120 to 140 meV \cite{kive03,birg06,head10}.  The optical excitation gap of 1.5--2~eV is of the charge-transfer type and does not depend on the presence of AF order  \cite{toku90}.  The hole on each Cu is in the $3d_{x^2-y^2}$ orbital \cite{tran87c,heal88,more11} with some hybridization to the neighboring O $2p_\sigma$ orbitals \cite{walt09}, where the degree of hybridization can vary among cuprate families \cite{rybi16}.  Other $3d$ levels are separated by at least 1.5~eV \cite{more11}.  When a modest density of holes is doped into the planes, the optical gap does not collapse; instead, weight is gradually transferred into the gap \cite{uchi91}.  The dopant-induced holes have a dominant O $2p_\sigma$ character \cite{chen91,bian88}.

\begin{figure}[t]
 \centering
    \includegraphics[width=0.9\columnwidth]{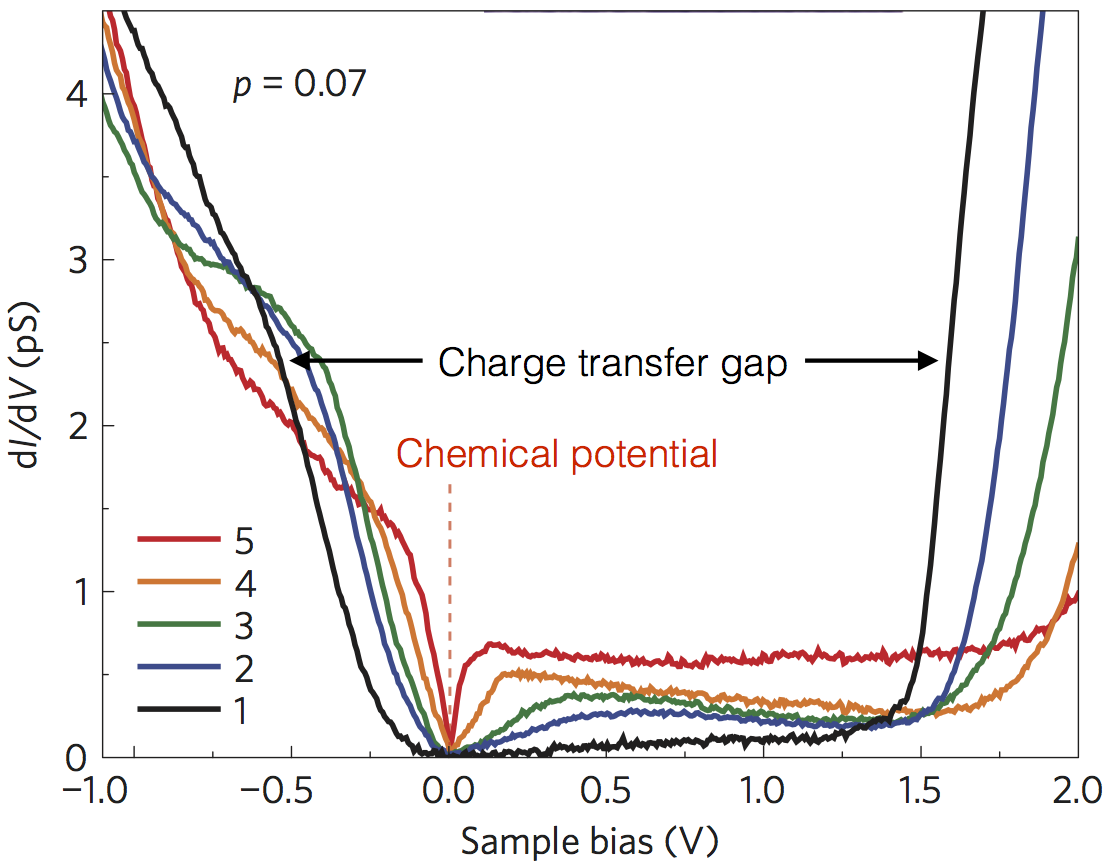}
    \caption{\label{fg:cai}  Tunneling curves at various points on the surface of Bi$_2$Sr$_{2-x}$La$_x$CuO$_{6+\delta}$ with $p = 0.07$ (charge-ordered insulator) measured by STM.   Reprinted with permission from \cite{cai16}, Springer Nature \copyright 2016. }
\end{figure}

Recent STM measurements on lightly doped Bi$_2$Sr$_{2-x}$La$_x$CuO$_{6+\delta}$ (Bi2201) confirm some of the earlier results and provide important new clues \cite{cai16}.  Figure~\ref{fg:cai} shows tunneling conductance measurements made at several different points within 200~\AA\ of one another on the same atomically-flat sample surface.  The hole concentration is estimated to be $p=0.07$, but the sample is not superconducting.  A crucial feature of these spectra is that the range of bias voltage spans the charge-transfer gap.  Looking at curve 1, we see that there are essentially no states within the gap at that point and the charge transfer gap is about 2~eV.  For the other curves, we see that, locally, weight from both the lower and upper edges of the gap has moved into the gap, and the chemical potential is pinned within the gap, not at the top of the valence band.  Furthermore, the transferred weight is spread over a large energy scale ($\sim0.5$~eV), and there is a gap centered at the chemical potential that is relatively particle-hole symmetric.

The variation in the local weight of the mid-gap states suggests a significant role for the long-range Coulomb interaction.  In the present case, a hole is introduced by replacing a La$^{3+}$ ion with a Sr$^{2+}$.  The hole goes into a neighboring CuO$_2$ plane, but it is not free to move away.  It feels a substantial Coulomb attraction to the dopant site, and in the limit of very small doping, the screening comes mainly from phonons \cite{kast98}.   As reviewed in \cite{kast98}, a single hole may be localized on a scale of a few lattice spacings.  In \lsco\ with a hole concentration $p\lesssim0.05$, the low-temperature in-plane resistivity shows a resistive upturn at low temperature with a form consistent with variable-range hopping in a Coulomb potential \cite{kast98,shkl84}.  

\begin{figure}[b]
 \centering
    \includegraphics[width=0.99\columnwidth]{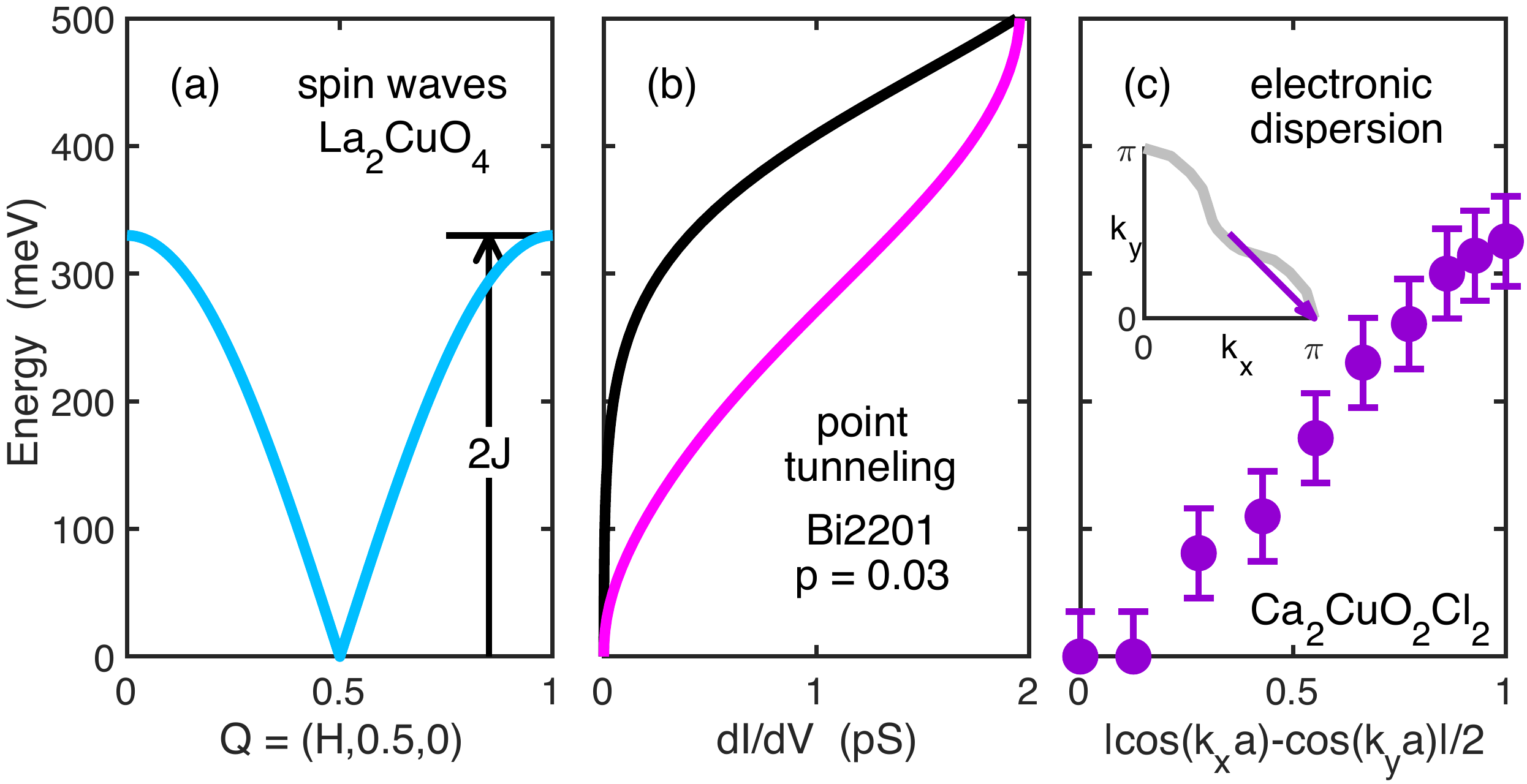}
    \caption{\label{fg:big}  (a) Spin-wave dispersion in La$_2$CuO$_4$ measured by neutron scattering \cite{cold01,head10}.  (b) Schematic version of typical conductance curves measured by STM at typical locations in Bi$_2$Sr$_{2-x}$La$_x$CuO$_{6+\delta}$ with $p = 0.03$ \cite{cai16}, plotted as binding energy vs.\ conductance; black: region with no in-gap states; magenta: region with significant in-gap states.  (c) Effective gap or dispersion along the nominal weak-coupling Fermi surface measured by ARPES in Ca$_2$CuO$_2$Cl$_2$ \cite{ronn98}. }
\end{figure}

\begin{figure*}[t]
 \centering
    \includegraphics[width=1.5\columnwidth]{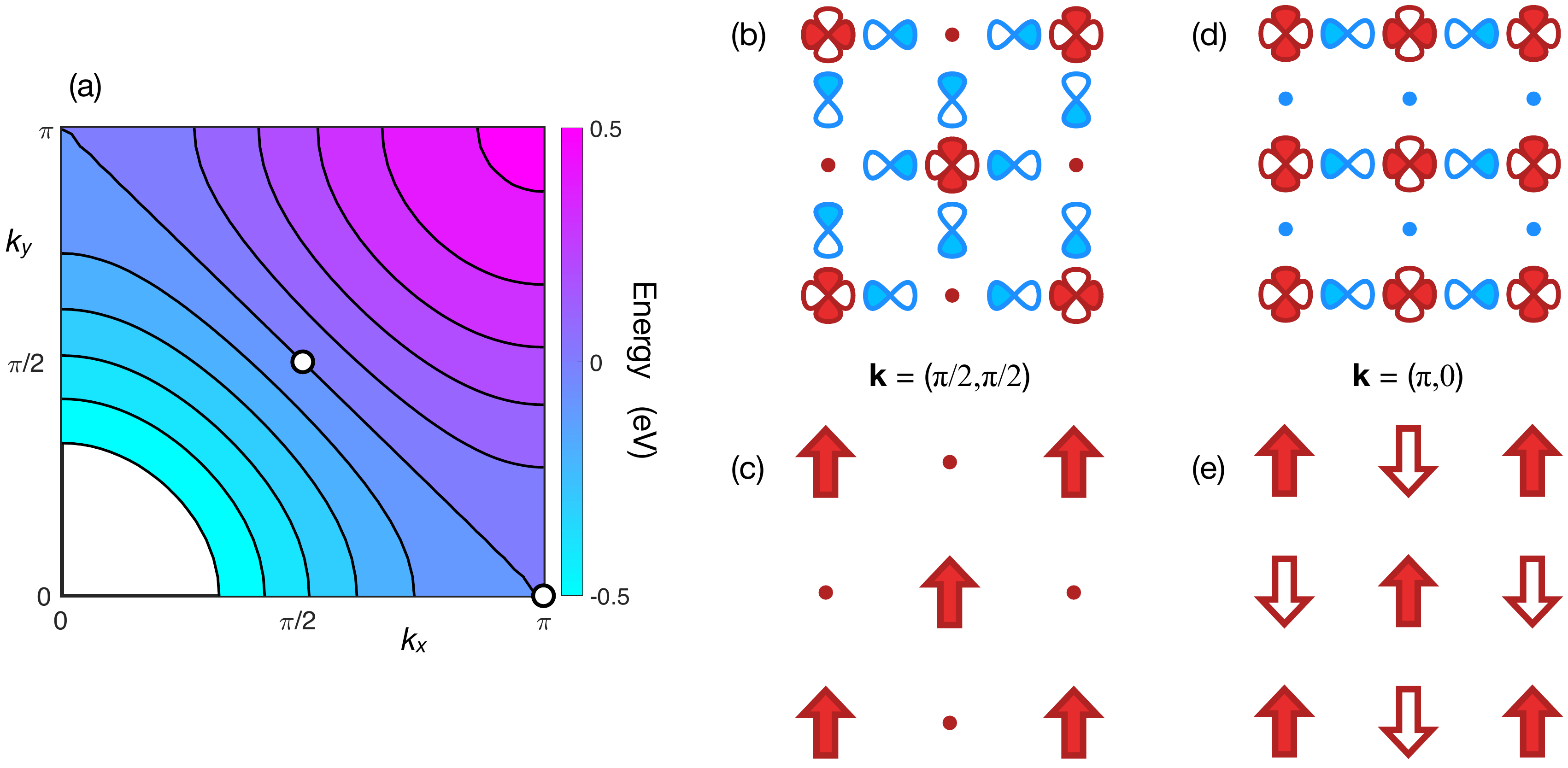}
    \caption{\label{fg:es}  (a) Dispersion of the antibonding band for a non-interacting 3-band model with only nearest-neighbor hopping, projected onto a quadrant of the first Brillouin zone with energy referenced to the chemical potential at half-filling.  Circles denote the points on the Fermi surface at $(\pi/2,\pi/2)$ and $(\pi,0)$.  (b) Schematic wave function for ${\bf k} = (\pi/2,\pi/2)$ \cite{ande95}.  (c) Antiferromagnetic spin structure with spin size proportional to the $d_{x^2-y^2}$ weights in the $(\pi/2,\pi/2)$ wave function.  (d) Schematic wave function at ${\bf k} = (\pi,0)$ \cite{ande95}, and (e) the corresponding weighted spin components. }
\end{figure*}

Coulomb effects are not the only impediment to charge motion; there is also the important role of AF correlations.  Here the relevant energy scale is the nearest-neighbor superexchange interaction $J$.  It can be determined from measurements of the antiferromagnetic spin waves in parent compounds such as La$_2$CuO$_4$ by inelastic neutron scattering, as illustrated in Fig.~\ref{fg:big}(a) \cite{cold01,head10}.  While longer-range interactions can have small impacts, one can see that the spin-wave bandwidth is $\sim 2J \sim 300$~meV.  This is also the energy scale for locally flipping a single spin.

It is interesting to compare the magnetic energies with the electronic dispersion measured by angle-resolved photoemission spectroscopy (ARPES) in another insulator, Ca$_2$CuO$_2$Cl$_2$, as shown in Fig.~\ref{fg:big}(c) \cite{ronn98}.  By knocking out an electron, a hole is created, whose energy varies with wave vector.  The measurements shown are roughly along the direction of the anticipated Fermi surface for noninteracting electrons (see Sec.~\ref{sc:rough}), in which case, the sample would be a metal and we would expect to see no dispersion.  Instead, the measured dispersion has a form quite similar to that of a $d$-wave superconducting gap, which is linear in $|\cos(k_xa)-\cos(k_ya)|$ \cite{ronn98}.  The magnitude of the effective gap here is very similar to $2J$, suggesting a connection to spin correlations.

To make a connection with the STM measurements, representative tunneling conductance curves for Bi$_2$Sr$_{2-x}$La$_x$CuO$_{6+\delta}$ with $p\sim0.03$ are shown in Fig.~\ref{fg:big}(b) \cite{cai16}.  Where in-gap states appear, their relative weight (on the positive binding-energy side) has a maximum that is comparable to $2J$.

\section{Rough interpretation}
\label{sc:rough}

To appreciate how the different results in Fig.~\ref{fg:big} may be connected, it is helpful to use simple ideas about the possible electronic structure.  We consider the tight-binding calculation involving the Cu $3d_{x^2-y^2}$ and O $2p_\sigma$ orbitals [where $\sigma = x (y)$ for O sites along the $a (b)$ axis] with only nearest-neighbor hopping.  (This is essentially the Emery model \cite{emer87}, but with the onsite Coulomb repulsion $U$ set to zero.)  The relevant formulas have been given in detail by Andersen {\it et al.}\ \cite{ande95}.  Using common parameter values (3~eV for the separation between the Cu and O levels and hopping $t_{pd}=1.6$~eV \cite{ande95}), the antibonding band, which should be half-filled for an undoped CuO$_2$ layer, is easily calculated.  The dispersion, projected onto a quadrant of the first Brillouin zone, is shown in Fig.~\ref{fg:es}(a).  In this simple model, the Fermi surface runs diagonally from $(\pi,0)$ to $(0,\pi)$ through $(\pi/2,\pi/2)$, the position of the node in the case of the $d$-wave superconducting gap (the nodal point).  Note that for this section the lattice parameter $a$ is set equal to 1.

The schematic wave functions at two of these points [indicated by circles in Fig.~\ref{fg:es}(a)] are also given in \cite{ande95}, and we reproduce them in Fig.~\ref{fg:es}(b) and (d).  The antibonding band involves arrangements of the O $2p_\sigma$ orbitals that are in-phase with the Cu $3d_{x^2-y^2}$.  At each wave vector, the variation of the wave function in real space follows $\cos({\bf k}\cdot {\bf r}_n)$, where ${\bf r}_n$ is the center of the $n^{\rm th}$ unit cell.  For ${\bf k} = (\pi/2,\pi/2)$, we see that the phase of the $d$ orbitals is identical for all sites along the diagonal, but is staggered for second neighbors along the Cu-O bond directions; for nearest neighbors, the amplitude is zero.  For ${\bf k}=(\pi,0)$, the $d$ orbitals are in-phase along [0,1] and antiphase along [1,0].  Again, these states have identical energies in this noninteracting model.

Now consider the impact of antiferromagnetism.  At the nodal point, the Cu sites that contribute to the wave function all have the same spin direction, as shown in Fig.~\ref{fg:es}(c), so that the electronic state is not impacted by the spin correlations.  On the other hand, at the antinodal wave vector $(\pi,0)$, neighboring Cu sites have antiparallel spins.  An extended state must have a single spin state, so the wave function at $(\pi,0)$ is incompatible with AF spin correlations.  Note that the incompatibility does not depend on static order; if the instantaneous spins are antiparallel on neighboring sites, then they cannot be part of the same extended state.  (For a proper analysis of the problem of one hole in a two-dimensional (2D) antiferromagnet, see \cite{lau11,ebra14}.)

In order to realize the state shown in Fig.~\ref{fg:es}(d), it is necessary to flip half of the Cu spins.  The energy to flip a spin corresponds to the maximum spin-wave energy, $\sim2J \sim 300$~meV.  Hence, it is plausible that, for a hole moving in an AF, the energy difference between the antinodal (AN) and nodal points is $\sim 2J$.  Furthermore, we can use the factor $\cos({\bf k}\cdot {\bf r}_n)$ to estimate the energy gap along the noninteracting Fermi surface in terms of the relative amplitude on neighboring Cu sites.  It has the same form as the absolute magnitude of the $d$-wave superconducting gap, consistent with the dispersion observed by ARPES in Fig.~\ref{fg:big}(c).

We can also use this picture to interpret the STM conductance curves.  In terms of density of states, the noninteracting band dispersion in Fig.~\ref{fg:es}(a) results in a van Hove singularity at the antinodal points.  AF correlations will smear out this distribution, but more importantly, they will put a gap in the middle of it.  Note that there are similar densities of states above and below the Fermi level, and the AF correlations will gap states on both sides of the chemical potential.  This is qualitatively compatible with the experimental observations, as in Fig.~\ref{fg:cai}.

Another factor for the STM concerns the nature of the tunneling process from the probe tip to the sample surface and along the $c$ axis, typically through an apical O site, to a Cu site in the CuO$_2$ plane nearest the surface.  The $2p$ orbital on the apical site has $s$ symmetry relative to the Cu, so that it cannot couple to the $3d_{x^2-y^2}$ but may couple to the in-plane O $2p_\sigma$ states.  From Fig.~\ref{fg:es}(b) and (d), one can see that no coupling is possible at ${\bf k}=(\pi/2,\pi/2)$ because the O orbitals are all in phase with the nearest Cu $3d_{x^2-y^2}$ orbital, but there is a finite coupling at $(\pi,0)$ as the phasing is different.  Such effects were originally noted in analyses of $c$-axis conduction and planar tunneling \cite{chak93,xian96}, and are discussed for STM in \cite{bala06}.

It is important to note that the competition between AF correlations and kinetic energy at the AN point does not depend on static order of the spins.  What matters is the strength of local, instantaneous correlations, and these can survive to very high temperatures \cite{imai93,birg99}, far beyond any ordering temperature.

\section{Large pseudogap}

Based on the comparison of magnetic and electronic spectra in undoped to very lightly hole-doped cuprates, it seems reasonable to associate the large pseudogap in the AN region with AF correlations.  In this section, we will follow how these energies change with doping and see how they are correlated.

Hole doping as low as $p=0.02$ is enough to destroy long-range AF order in cuprates \cite{birg06}.  At low excitation energies, the long-wavelength spin correlations are significantly reorganized \cite{enok13}; however, the high-energy excitations evolve more gradually \cite{stoc10,fuji12a}.  A useful measure of the high-energy evolution is given by Raman scattering measurements of two-magnon excitations \cite{suga03}.  The excitation mechanism for a quasi-2D AF involves flipping the orientation of a nearest-neighbor pair of spins with respect to all others.  This disrupts 6 magnetic bonds and costs an energy of $\sim3J$ \cite{chel18}.  

\begin{figure}[b]
 \centering
    \includegraphics[width=0.8\columnwidth]{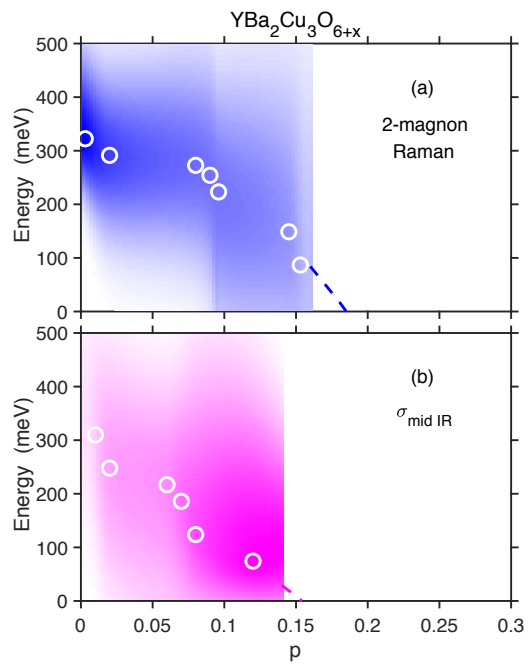}
    \caption{\label{fg:mYBCO}  (a) Two-magnon response measured by Raman scattering in \ybco\ as a function of estimated hole concentration $p$ \cite{suga03}. (b) Mid-infrared optical conductivity in \ybco\ \cite{lee05}, with the energy scale divided by 2.  In both (a) and (b), the data have been interpolated, and circles denote measured peak positions.  Dashed lines are quadratic extrapolations of the peak positions.}
\end{figure}

Figure~\ref{fg:mYBCO}(a) shows how the two-magnon scattering evolves with doping in \ybco\ \cite{suga03}.  As one can see, the well-defined peak in the lightly-doped system softens in energy as holes are added.  Sugai {\it et al.} \cite{suga03} have similar results for  \lsco\ (LSCO), \bscco\ (Bi2212), and Bi2201, extending to higher doping, and in all cases the 2-magnon peak appears to become overdamped in the vicinity of $p\sim0.2$.  If we assume that the excitation mechanism remains the same in the doped systems, then the energy will depend both on the strength of $J$ and the degree to which a finite patch of Cu sites retains AF correlations.  While the AF correlations restrict the hole motion and spatial distribution, the increasing density of holes must reduce the spatial areas in which AF correlations among Cu spins can survive.  Neutron scattering measurements have shown that the instantaneous spin correlation length is reduced to approximately one lattice spacing near optimal doping \cite{xu07}, so a decrease in correlation length can describe much of the decay in the peak energy with $p$.  Nevertheless, it can also be convenient to think of an average reduction in $J$.  For example, Johnston \cite{john89} pointed out that the temperature dependence of the bulk magnetic susceptibility in \lsco\ can be scaled by doping-dependent values of $J$ and average moment per Cu, with both trending toward zero at $p=x\sim0.2$, as confirmed by others \cite{naka94}.

I argued in the last section that the energy scale of the distribution of mid-gap states detected by STM in lightly doped cuprates is associated with AF correlations.  Another measure of these states is given by optical conductivity, $\sigma_1(\omega)$.  At energies below that of the charge-transfer gap, optical measurements largely probe transitions from filled to empty mid-gap states.  For moderately underdoped cuprates, it is possible at low temperature to distinguish the broad ``mid-IR" conductivity from the more coherent Drude peak, centered at zero energy \cite{lee05}.  Figure~\ref{fg:mYBCO}(b) shows a rough version of the mid-IR feature observed in YBCO at temperatures close to but above the superconducting $T_c$ \cite{lee05}.  Here the energy scale has been divided by two to make it comparable to electron binding energies.   The energy scale decreases in a fashion similar to the two-magnon energies, while the spectral weight grows with doping.

\begin{figure}[t]
 \centering
    \includegraphics[width=0.6\columnwidth]{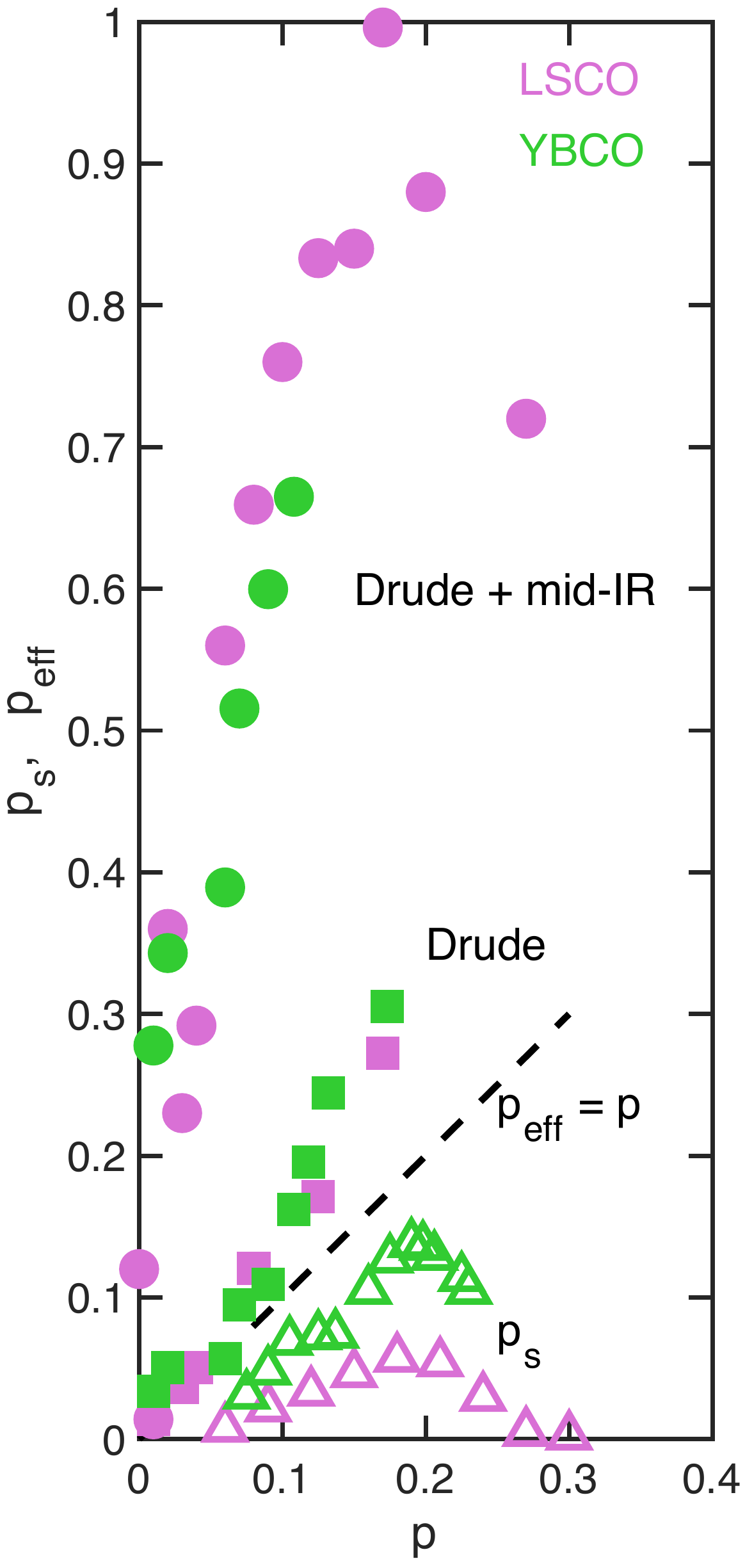}
    \caption{\label{fg:p} Comparison of various estimates of carrier density in \lsco\ (violet) and \ybco (green).  Triangles: superfluid density, $p_s$, from measurements of the magnetic penetration depth on Y$_{0.8}$Ca$_{0.2}$Ba$_2$Cu$_3$O$_{6+x}$ by muon spin rotation \cite{bern01} and on LSCO films by mutual inductance \cite{lemb11}; squares: effective carrier density, $p_{\rm eff}$, from integrating in-plane optical conductivity to 80 meV \cite{padi05}; circles: $p_{\rm eff}$ from integrating to 1.5 eV \cite{padi05,uchi91}.  In evaluating $p_{\rm eff}$, the effective masses of $4m_e$ and $3m_e$ for LSCO and YBCO, respectively, determined in \cite{padi05} were used. }
\end{figure}

The variation of the effective carrier concentration is considered more quantitatively in Fig.~\ref{fg:p}.  Integrating the optical conductivity from zero to a finite frequency gives a result proportional to the effective hole density $p_{\rm eff}$ divided by an effective mass $m^*$.  Padilla {\it et al.}\ \cite{padi05} compared optical data with measures of $p$ from the Hall coefficient to obtain $m^*/m_e\approx4$ for LSCO and 3 for YBCO, where $m_e$ is the electron mass.  (A later study has suggested that $m^*$ decreases by $\sim50$\%\ with doping in the range $0.1 <p<0.22$ \cite{vanh09}.)  To estimate the carrier density associated with the Drude peak, they integrated $\sigma_1(\omega)$ data for LSCO and YBCO to 80 meV, which yields the $p_{\rm eff}$ values indicated by filled squares in Fig.~\ref{fg:p}, plotted against the estimated dopant-induced carrier density $p$ \cite{padi05,lee05}.  At small $p$, the effective Drude carrier density is close to $p$, while it begins to rise above $p$ for $p\gtrsim0.1$; such behavior in LSCO was originally noted based on measurements of the Hall coefficient \cite{taka89b}, and similar behavior is seen in the nodal weight detected by ARPES \cite{yosh03}. Integrating up to 1.5~eV, one captures the $p_{\rm eff}$ associated with both the Drude peak and the mid-IR signal, indicated by the filled circles.  This value is several times larger than the Drude weight alone, and approaches 1 for LSCO at $p\approx0.18$; however, it never seems to reach the level of $1+p$ that one would expect in the absence of correlations.

In a complementary fashion, one can obtain the superfluid density $p_s$ from measurements of the magnetic penetration depth $\lambda$ at $T\ll T_c$, using \cite{tink75} 
\begin{equation}
\frac{1}{\lambda^2} = \frac{4\pi p_se^2}{m_ec^2}.
\end{equation}
Two ways to measure $\lambda$ are by muon spin relaxation ($\mu$SR) in an applied magnetic field \cite{uemu89} and by mutual inductance \cite{lemb11}.  The open triangles in Fig.~\ref{fg:p} indicate $p_s$ values for Y$_{0.8}$Ca$_{0.2}$Ba$_2$Cu$_3$O$_{6+x}$ (determined by $\mu$SR \cite{bern01}) and for LSCO thin films (from mutual inductance \cite{lemb11}).  (A related figure of $p_s$ in LSCO, including detailed data from \cite{bozo16}, is presented in \cite{pelc19}.)  Much of the Drude weight from $T>T_c$ goes into the superfluid, while $p_s$ in these materials reaches a maximum near $p\sim0.2$.

\begin{figure}[t]
 \centering
    \includegraphics[width=0.8\columnwidth]{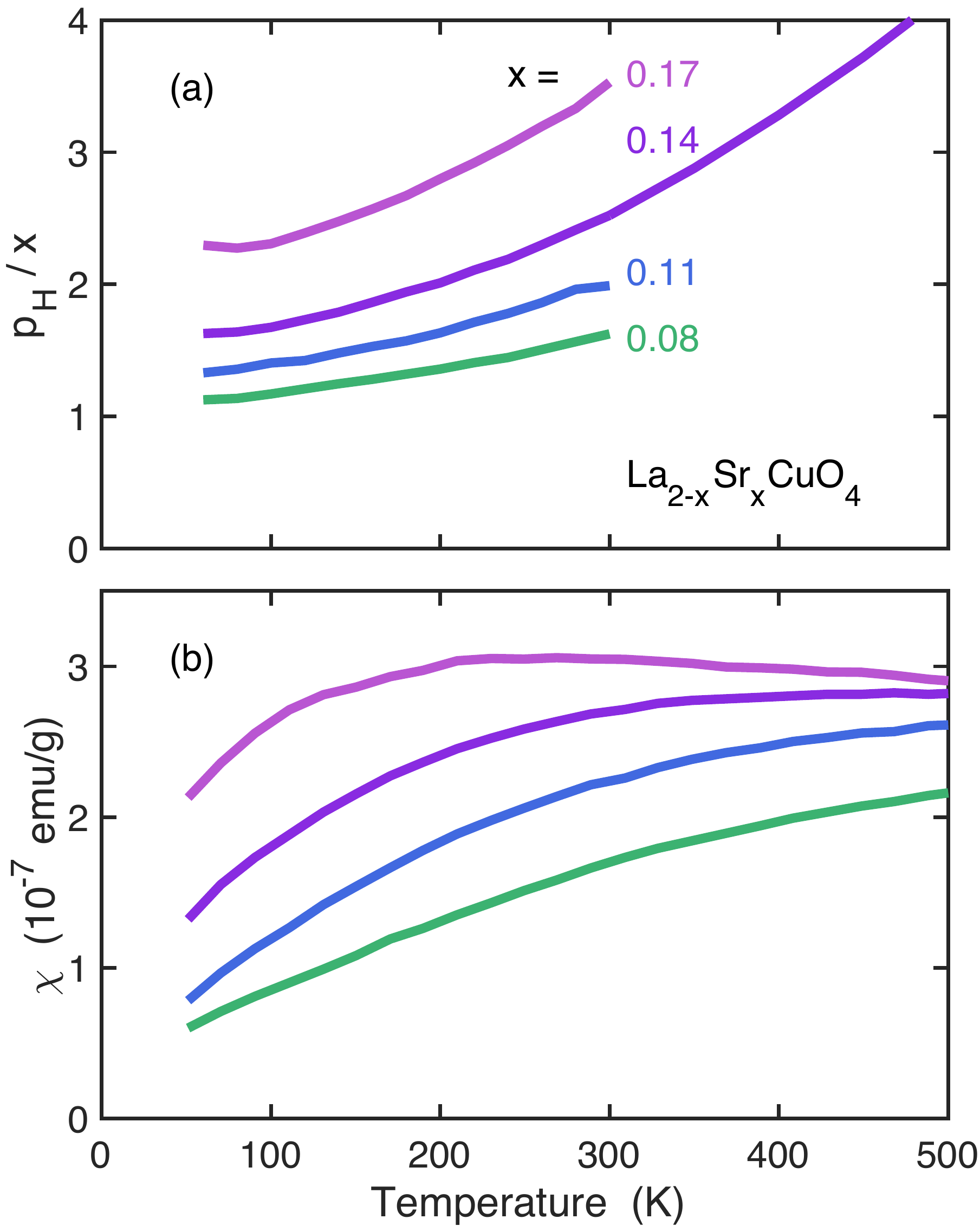}
    \caption{\label{fg:Hallchi}  (a) Ratio of the hole concentration $p_{\rm H}$ determined from the Hall effect to the Sr fraction $x$ in \lsco; data from \cite{ando04} have been interpolated. (b) Bulk magnetic susceptibility in \lsco; data from \cite{naka94} have been interpolated.}
\end{figure}

The hole concentration $p_{\rm H}$ measured by the Hall effect shows temperature dependence.  Figure~\ref{fg:Hallchi}(a) shows the the ratio $p_{\rm H}/x$ for \lsco\ from measurements reported in \cite{ando04}.  The ratio grows both with temperature and doping.  This behavior is correlated with changes in the magnetic susceptibility, shown in Fig.~\ref{fg:Hallchi}(b) \cite{naka94}.  The magnetic susceptibility for a purely magnetic system is inversely proportional to $J$; it is small when AF correlations are strong, and it rises with temperature to a maximum as nearest-neighbor correlations become weak.  The maximum in $\chi$ has been used as one measure of the pseudogap, and the correlation with the temperature dependence of $p_{\rm H}$ has been analyzed previously \cite{hwan94,gork06}.  Optical studies of underdoped cuprates \cite{take02b,take03,lee05} have shown that the electronic excitations are completely incoherent at high temperature, where the AF correlations are also weak; a Drude peak only develops on cooling, which occurs together with the enhancement of the spin correlations (also seen by inelastic neutron scattering \cite{aepp97}).  

The large pseudogap effectively closes near $p_c\sim0.2$ \cite{tall01} as the fraction of correlated Cu spins is reduced sufficiently that they can be overdamped by quasiparticle scattering in the normal state.  With the loss of dominant magnetic correlations, the antinodal states can become more conventional.  It is in the vicinity of $p_c$ that quasiparticle interference analysis of STM spectroscopic images observes a transition of the locus of Bogoliubov states from finite arcs to a full Fermi surface \cite{fuji14a}.  In LSCO, it is close to $p_c$ that the change in Fermi-surface topology from hole-like to electron-like is observed \cite{yosh07}.

\section{Small pseudogap}

From the earliest spectroscopic imaging with STM on Bi2212 at $T\ll T_c$, a common point of emphasis has been the spatial variation and disorder in apparent superconducting coherence peaks with energies in the range of 20--50~meV \cite{cren00,howa01,pan01}; such behavior is also seen in Bi2201 \cite{boye07}.  When measurements were eventually done across $T_c$, it was found that there is also a correlation between the size of the local gap and the temperature at which it closes, with larger gaps closing at higher temperatures that extend well above the bulk $T_c$ \cite{gome07,pasu08}.   Experiments also established correlations between dopant defects and local gap size \cite{mcel05b,zelj12}.

The scale of the disorder in the coherence peaks of Bi2212 is tens of \AA\ \cite{fang06}, but this is also the scale of the superconducting coherence length \cite{wang03b}.  This has led to proposals that the inhomogeneity is associated with variations in the local hole concentration and the local pairing scale \cite{wang01,fang06}.   There is also evidence of a spatial variation of hole concentration in \lsco\ based on the analysis of $^{63}$Cu nuclear quadrupole resonance \cite{sing02a} and $^{17}$O nuclear magnetic resonance \cite{sing05} measurements.  For $x\sim0.15$, the width of the distribution is $\Delta x\sim 0.05$ at low temperature, with a characteristic length scale of $\sim30$~\AA.  The lack of effective charge screening is also indicated by a study of the dynamic charge susceptibility with momentum-resolved electron-energy-loss spectroscopy in Bi2212, where the expected plasmon features are missing \cite{mitr18}.

\begin{figure}[b]
 \centering
    \includegraphics[width=0.99\columnwidth]{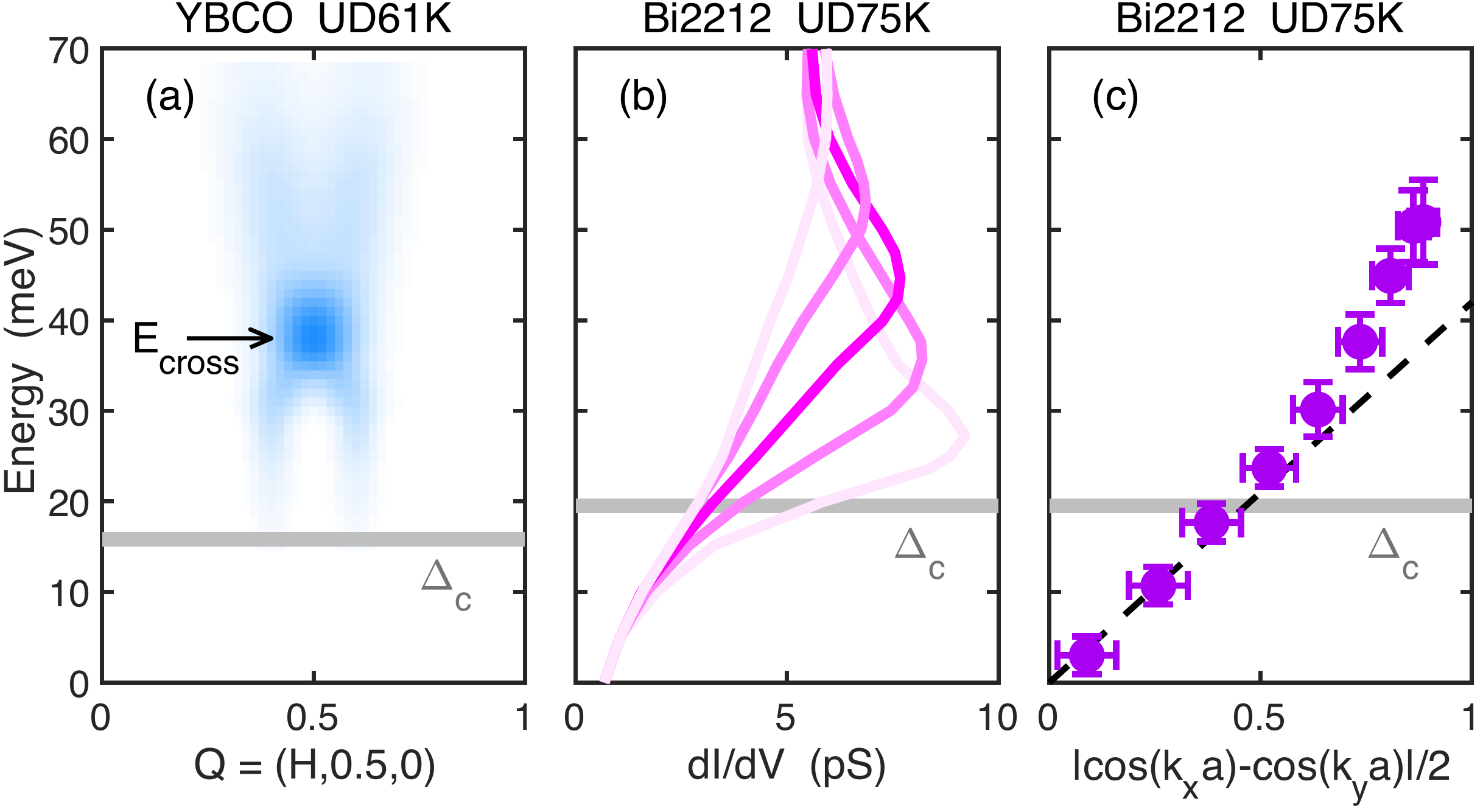}
    \caption{\label{fg:small}  (a) Magnetic spectral weight measured by neutron scattering in YBa$_2$Cu$_3$O$_{6.6}$ with $T_c=61$~K and $p\approx0.12$ \cite{hink04,hink07}.  (b) Bias voltage (binding energy) vs.\ conductance measured by STM for typical regions in Bi2212 with $T_c=75$~K and $p = 0.13(1)$ \cite{mcel05a}; shading reflects the relative frequency with which these spectra occur.  (c) Superconducting gap dispersion measured by ARPES at 10~K in a similar underdoped Bi2212 sample with $T_c=75$~K \cite{lee07}; dashed line indicates the simple $d$-wave gap form.  In each panel, the gray line indicates $\Delta_c=3kT_c$ based on Raman results \cite{sacu13,munn11}. }
\end{figure}

Figure~\ref{fg:small}(b) shows typical low-temperature conductance curves (for positive binding energy) from STM measurements at various locations on the surface of underdoped Bi2212 \cite{mcel05a}.  For a sample with $T_c=75$~K, the most frequently observed gap energy is 45~meV, with a broad spread of gaps observed around that value.  (Note that for more underdoped samples, the energy scale of the coherence peaks decreases and is clearly distinguished from the large pseudogap \cite{kohs07,ruan18}.)

Another measure of the superconducting gap is given by ARPES, which provided some of the original evidence that the gap has a $d$-wave form \cite{dama03}.   Deviations from the $d$-wave gap dispersion are observed in underdoped cuprates \cite{kond09,vish12}, as illustrated in Fig.~\ref{fg:small}(c) for Bi2212 with $T_c=75$~K \cite{lee07}.  The variation from the $d$-wave-like dispersion indicated by the dashed line occurs in the antinodal region and at energies that are similar to the coherence-peak energies detected by STM.  As ARPES averages over a large area, it is reasonable to expect that it averages over the distribution of local gaps detected by STM.  This averaging effect was recently taken into account to explain the anomalous temperature dependence near $T_c$ of the antinodal spectral function in slightly-overdoped Bi2212 crystals \cite{zaki17}.  It should also explain the appearance of sharp antinodal peaks at $T\gtrsim T_c$ in optimally to underdoped Bi2212 \cite{fedo99,ding01}.  

On warming to $T\sim T_c$, ARPES measurements on underdoped Bi2212 and Bi2201 find that the gap closes only along a finite arc, centered on the nodal point, with a gap remaining in the AN region \cite{norm98,lee07,kond09,kond13,kond15}.  The magnitude of the low-temperature gap that develops at the ends of the arcs has been labelled $\Delta_c$, the coherent superconducting gap, which is detected by Andreev reflection in tunneling spectroscopy \cite{deut99} and by Raman scattering \cite{deve07}.  For the families YBCO, Bi2212, Tl$_2$Ba$_2$CuO$_{6+\delta}$, and HgBa$_2$CuO$_{4+\delta}$, it has been observed that $2\Delta_c\approx 6kT_c$ \cite{munn11,sacu13}.  For energies below $\Delta_c$, STM studies observe spatially homogeneous behavior \cite{push09,kuro10,sugi17}.  The magnitude of $\Delta_c$ is plotted as a gray horizontal line in Fig.~\ref{fg:small}.  As one can see, the variations among the tunneling curves tend to disappear below $\Delta_c$ and the ARPES gap follows the $d$-wave slope below that energy.

The energy scale associated with the antinodal coherence peaks is smaller than that of the big pseudogap, and hence it seems reasonable to label it the small pseudogap.  As with the big pseudogap, it is appropriate to consider the connection with antiferromagnetic correlations.  Figure~\ref{fg:small}(a) shows a schematic version of the imaginary part of the dynamic susceptibility, $\chi''({\bf q},\omega)$, in YBa$_2$Cu$_3$O$_{6.6}$ as determined by inelastic neutron scattering \cite{hink07,hink10}.  (No neutron scattering studies of underdoped Bi2212 crystals have been reported yet.) One sees an ``hourglass'' spectrum similar to that of other underdoped cuprates \cite{fuji12a}, with the neck of the hourglass at $E_{\rm cross} =38$~meV.  The excitations dispersing above $E_{\rm cross}$ are similar to spin waves with a substantial gap.  Such a singlet-triplet gap is generally considered by theorists to be good for inducing pairing among the doped holes \cite{dahm18,dago96,scal12a}.  

At the same time, downwardly-dispersing excitations are also present, and these are predicted to be bad for pairing \cite{dahm18}.  The magnitude of $\Delta_c$ corresponds to the scale where the low-energy excitations are getting very weak.  A recent study has found that this is a general feature: the spin gap $\Delta_{\rm spin}$, defined by the energy below which magnetic spectral weight decreases below $T_c$, acts as an upper limit to $\Delta_c$ \cite{li18}.  When combined with the evidence for charge inhomogeneity, I believe that this provides an important clue to the physics behind $\Delta_c$.  The AF excitations extend to lower energies for smaller hole concentrations.  Defects, such as Zn impurities, can also locally-induce low-energy AF excitations \cite{kaku93,such10,maha94,juli00}.  From the STM imaging results, we have have a picture of patches with local superconducting coherence that can develop at temperatures above the bulk $T_c$.  One can think of an analogy with granular superconductors, where individual grains develop strong pairing but bulk superconductivity is limited by the lack of phase coherence across the interfaces between grains \cite{sace11,prac16}.  In fact, this analogy has been invoked previously \cite{imry12} to explain the correlation between superfluid density and the product of $T_c$ with normal-state conductivity, known as Homes' law \cite{home04b} and which subsumes the Uemura relation between $T_c$ and superfluid density in the underdoped regime \cite{uemu89}.  Furthermore, the temperature-dependent development of a coherent superconducting gap $\Delta_c$ smaller than the superconducting pseudogap in a granular superconductor has been demonstrated recently by Andreev spectroscopy on indium oxide samples \cite{dubo19}.  For cuprates, the barrier to phase coherence would be regions with low-energy AF correlations.  Developing superconducting phase order across the sample requires a proximity-effect-like establishment of a minimum spin gap in all spatial regions.  

Several analyses of ARPES data on Bi2212 conclude that the closing of the superconducting gap along a finite arc is associated with the electronic scattering rate exceeding the gap size in the near nodal region \cite{norm07b,rame11,rebe12}.  A study of $c$-axis optical conductivity found evidence for superconducting phase fluctuations in the vicinity of $T_c$ \cite{cors99}, and further support was provided by an STM study of quasiparticle interference as a function of temperature in a fairly underdoped Bi2212 sample \cite{lee09}.  Recent observations of nonlinear conductivity extending well above $T_c$ in various cuprates have been interpreted as evidence of charge inhomogeneity \cite{pelc18}.  Together, these results seem compatible with and supportive of the proposed scenario.

\begin{figure}[t]
 \centering
    \includegraphics[width=0.9\columnwidth]{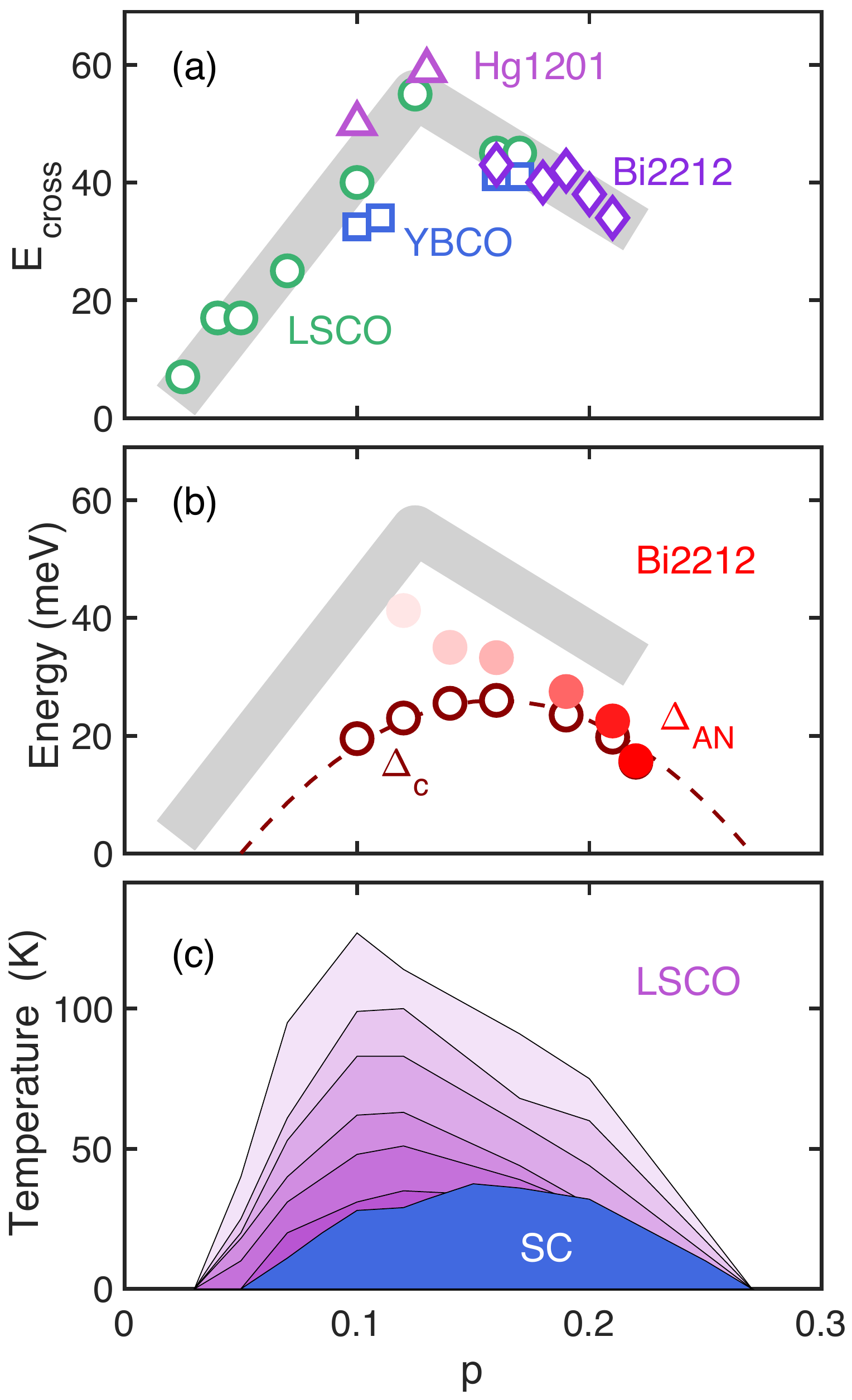}
    \caption{\label{fg:small_dope}  (a) $E_{\rm cross}$ from neutron scattering studies of \lsco\ \cite{fuji12a}, \ybco \cite{stoc05,hayd04,rezn08,woo06}, \bscco \cite{fong99,xu09,fauq07,he01,capo07}, and HgCa$_2$CuO$_{6+\delta}$ \cite{chan16b,chan16c}.  Gray line indicates the common trend among these families.  (b) Antinodal gap (red filled circles) and coherent gap (dark red open circles) energies from Raman scattering on \bscco\ \cite{blan10,sacu13}.  The shading of the $\Delta_{\rm AN}$ symbols reflects the change in peak area with doping. (c) Violet contours indicate relative strength (logarithmic scale) of superconducting fluctuations above the superconducting phase (blue) obtained from Nernst effect measurements on \lsco\ \cite{wang06}. }
\end{figure}

To test the correlation between $E_{\rm cross}$ and the AN SC gap energies, it is useful to look at the doping dependence of these quantities.  Figure~\ref{fg:small_dope}(a) shows the doping dependence of $E_{\rm cross}$ determined by neutron scattering in four families of cuprates.  The trends are remarkably similar in all of these compounds.  The general doping trend is approximated by the gray bars in the background. 

In Fig.~\ref{fg:small_dope}(b), the gray bars are repeated to allow a comparison with measurements of $\Delta_c$ and the small AN pseudogap $\Delta_{\rm AN}$ determined by Raman scattering in Bi2212 \cite{blan10,sacu13}.  (Comparisons of $2\Delta_{\rm AN}$ from a broad range of techniques have been presented elsewhere \cite{hufn08}; the other measures are consistent with the Raman results.  For Raman results on a broader range of families, see \cite{munn11}.)   Besides indicating the energies by symbol position, the shading of the symbols is indicative of the integrated intensity of the spectral feature, which extrapolates to zero at $p=0.10$ \cite{sacu13}.  Note that the absence of the $\Delta_{\rm AN}$ signal for $p\lesssim0.12$ is found for other cuprate families, as well \cite{munn11}.  Also, ARPES measurements on Bi2212 with $p\lesssim0.1$ ($T_c < 70$~K) generally show an absence of legitimate quasiparticle peaks in the AN region, where the large pseudogap tends to dominate \cite{tana06,hash14,droz18,zhon18b}.  Hence, it appears that $\Delta_{\rm AN}\lesssim E_{\rm cross}$ (with the exception of LSCO, where $\Delta_{\rm AN}\ll E_{\rm cross}$ \cite{li18}).  This result is consistent with the idea that the singlet-triplet excitation gap for local Cu moments sets the scale for pairing.

Figure~\ref{fg:small_dope}(c) shows the temperature dependence of superconducting fluctuations as a function of doping in LSCO, where the measure is the Nernst coefficient, plotted with roughly logarithmic intensity contours \cite{wang06}.  (The Nernst effect is the transverse voltage measured in response to a longitudinal temperature gradient in the presence of a magnetic field; it is sensitive to vortex fluctuations for temperatures near $T_c$ \cite{wang06}.)  The maximum onset temperature occurs for $p\sim0.1$, with the response falling rapidly at lower $p$.  This behavior is consistent with the result in Fig.~\ref{fg:small_dope}(b) that the maximum local pairing gap is optimized near $p\sim 0.12$.  In contrast, phase coherence and superfluid density are optimized at $p\sim p_c$, as indicated by the plot of the superfluid density in Fig.~\ref{fg:p}.

\section{Discussion}

The interpretations presented above have implications for various other features observed in cuprates, as discussed below.

\subsection{Case of $\Delta_{\rm spin}=0$}

If the energies associated with superexchange among Cu spins and the delocalization of doped holes are competing, then one must eventually confront the issue of how correlated local moments and mobile holes are instantaneously distributed in real space.  This topic is the subject of intertwined orders, which has been reviewed elsewhere \cite{frad15}.  The evolution of the low-energy spin excitations, as a modest density of holes is introduced, shows a common shift from commensurate to incommensurate, with the incommensurability growing in proportion to $p$, for several cuprate families \cite{enok13}.  This behavior is simply connected to the rise of $E_{\rm cross}$ with doping \cite{fuji12a}.  Recent  numerical calculations support the idea that hole-doping an antiferromagnet leads to segregation of charge and spins into stripe-like structures \cite{huan17,zhen17}.

In the extreme experimental case, one observes static charge and spin stripe order, as in \lbco\ \cite{huck11}.  In the presence of maximum spin order at $x=1/8$, we have $\Delta_{\rm spin}=0$ and a strong suppression of spatially-uniform $d$-wave superconductivity, as indicated by the severely depressed onset temperature for 3D superconductivity.  Nevertheless, the occurrence of 2D superconductivity \cite{li07} can be rationalized in terms of a putative pair-density-wave superconductor \cite{berg09b,agte19}.  This can be viewed as the limiting case of low-energy spin fluctuations inhibiting uniform phase coherence between hole-rich regions with a large pairing scale.  Once spin order develops, the neighboring stripes of pair correlations can establish a relative $\pi$ phase shift to minimize the overlap with the spin stripes.  These are variations on a common theme.

\subsection{Charge order}

Charge-density-wave order has now been observed in most cuprate families, where, in contrast to \lbco, it develops only in the absence of spin order \cite{comi16}.  In particular, a significant range of doping has been tested in YBCO \cite{huck14,blan14}.  In zero-field, the onset temperature (which is not sharply defined) has a maximum of $\sim150$~K at $p\sim0.12$, with finite order observed for $0.08 \lesssim p\lesssim0.18$; the short-range and static characters are confirmed by NMR \cite{wu15}.  This ordering correlates with the onset of the spin gap as determined by the temperature dependence of the spin-lattice relaxation rate measured by Cu NMR \cite{taki91,baek12}.  Enhanced $\Delta_{\rm AN}$ peaks were found in this regime by Raman scattering \cite{suga03} and ARPES \cite{naka09}; in both cases, the amplitudes decay towards zero as $p$ decreases to 0.1.

Further evidence of superconducting correlations above $T_c$ is provided by studies of $c$-axis optical conductivity.  A transverse bilayer plasmon associated with interlayer Josephson coupling \cite{vand96} has been identified in YBCO \cite{grun00}.  In underdoped samples, it starts to develop well above $T_c$ \cite{home95,schu95,lafo09,dubr11}.  At low temperature, the connection with superconductivity is indicated by a partial suppression in a $c$-axis magnetic field \cite{lafo09}.  There is also evidence for a small amount of superfluid density that survives above $T_c$ in the same regime as the CDW order \cite{uyku14}.

There is disorder in this system associated with imperfect Cu-O chain ordering.   X-ray scattering studies of chain order in high-purity, underdoped samples find finite correlation lengths along all directions \cite{schl95,zimm03}.  Even the best case of a carefully annealed and  detwinned crystal had a correlation length $\xi_c$ of just 5 unit cells along the $c$ axis, while in-plane, $\xi_b\sim400$~\AA\ parallel to the chains and $\xi_a\sim150$~\AA\ in the perpendicular direction \cite{lian00}.  More typical values are $\xi_b<200$~\AA\ parallel to the chains \cite{schl95}. The finite correlation lengths have been confirmed by measurements in which a small x-ray beam is scanned across a sample \cite{ricc13,ricc14}.   NMR on chain-site Cu confirms the disorder through the presence of inequivalent sites \cite{wu16}.  The disorder appears to be reflected in the CDW measurements, where the typical in-plane correlation length is $\sim60$~\AA\ \cite{huck14,blan14}.

It seems likely that all of these phenomena are connected.  The evidence for local pairing correlations in a landscape of finite disorder could lead to a spatial modulation of the pair density.  Indeed, it has been proposed that local PDW order underlies the CDW correlations \cite{dai18}, in which case the small pseudogap might be associated with PDW order.

\subsection{Quantum oscillations}

Detailed studies of quantum oscillations (QO) have been performed on YBCO at high magnetic fields \cite{doir07,seba15}.  It is clear that the magnitude of the QO will be limited by impurities and disorder \cite{chak08}.  Is the observation of QO consistent with the disorder that has been observed in x-ray scattering studies?  To answer this, we must consider the length scales involved.

Finite QO gradually appear at fields greater than 20 T.  The most relevant quantity to consider would be the cyclotron radius, but this requires quantitative knowledge of the relevant Fermi velocity.  As an alternative, one can evaluate the vortex lattice spacing at the onset of QO.  

Assuming a square vortex lattice \cite{meli08,leos15}, the vortex spacing at 20 T is 144~\AA.  This is comparable to the in-plane correlation length along the chain direction \cite{schl95}.  With increasing field, the magnetic length decreases while the disorder remains fixed, so that the relative volume compatible with coherent cyclotron orbits will increase with field.  

\subsection{Tri-layer cuprates}

Within a given family, $T_c$ varies with the number of consecutive CuO$_2$ layers, showing a maximum for three layers \cite{chak04}.  Analysis of NMR Knight shift data on 3- and 4-layer cuprates indicates that the hole concentrations are different on the inner and outer layers, with $p$ being larger on the outer planes by $\sim0.04$ \cite{shim11}.  The results shown in Figs.~\ref{fg:p} and \ref{fg:small_dope} suggest why the distinct values of $p$ could be beneficial for $T_c$.  The superfluid density reaches a maximum for $p\sim0.2$, whereas pairing strength appears to be optimized for $p\sim0.12$.  The trilayer cuprates can simultaneously benefit from both of these features, consistent with a proposed mechanism for enhancing $T_c$ \cite{kive02b}.

\subsection{Other measures of disorder and SC fluctuations}

One must acknowledge that not all researchers are in agreement on the magnitude and relevance of disorder.  For example, there has been an analysis of charge disorder in LSCO, based on an analysis of x-ray diffraction peak widths and superconducting transition widths, which concludes that the range of superconducting fluctuations above $T_c$ is no more than 10\%\ of $T_c$ for a significant range of doping \cite{mosq09}.  There have also been studies of magnetoresistance in YBCO \cite{rull11} and in-plane optical conductivity on LSCO \cite{bilb11a} from which it was concluded that superconducting fluctuations are restricted to a limited temperature range above $T_c$.

These studies demonstrate that, to the extent that superconducting correlations survive well above $T_c$, they do not occur uniformly across a sample.  That observation can still be consistent with local patches that have a high transition temperature, provided that these patches do not percolate until the temperature gets close to $T_c$.  Regarding disorder, one must give significant weight to local probe studies.  For example, a scanning probe study of an LSCO film ($x\approx0.1$, $T_c\approx 18$~K) found a disordered distribution of diamagnetism that was detectable to $T\sim3T_c$ \cite{iguc01}.

\subsection{Evidence for QCP}

There are certainly interesting behaviors such as nematic order \cite{lawl10,nie14,mouk19} that evolve with doping in cuprates, and evidence for a possible nematic QCP has recently been presented \cite{auvr19}.  I view such order as more of a consequence than a driver of the electronic correlations and thus compatible with the perspective presented in this paper.

An analysis of specific heat measurements on Y$_{0.8}$Ca$_{0.2}$Ba$_2$Cu$_3$O$_{6+x}$ found that the superconducting condensation energy gained below $T_c$ shows a sharp peak at $p=p_c$ \cite{lora98,tall01}.  Of course, this analysis ignores pair condensation energy that may develop above $T_c$, which one expects to be more significant for $p< p_c$.  It is also relevant to note that the superfluid density for the same system varies much more gradually with $p$, as shown in Fig.~\ref{fg:p}.

 A recent study of the low-temperature electronic heat capacity in Nd- and Eu-doped LSCO, with superconductivity suppressed by a magnetic field of up to 18~T, finds a sharp peak at $p_c = 0.23$ \cite{mich19}, which is interpreted as evidence of a QCP.  Given that there are a structural transition in that vicinity \cite{buch94a} and large rare-earth moments that can amplify the correlations of Cu moments, there are reasons to wonder whether the spike in specific heat divided by $T$ ($C/T$) is purely electronic.  Perhaps a more interesting question is related to the large dip in $C/T$ found at $p\sim0.12$ \cite{mich19}.  The assumption has been made that the applied magnetic field is sufficient to destroy all electronic pairing; however, a recent high-field transport study on LBCO with $x=1/8$ finds 2D superconductivity at $T=0.35$~K and a field of 20~T, while there is evidence that pairing correlations may continue to survive at 35~T \cite{li19}.  Another way to destroy pairing is to substitute Zn for Cu.  A study of the electronic $C/T$ in LSCO, with normal-state behavior determined by Zn doping, found that the dip largely filled in and the maximum at $p\sim0.2$ was fairly rounded \cite{momo94}.

\section{Summary}

A broad range of experimental characterizations of cuprates can be reconciled with a few simple concepts.  There is a large pseudogap that results from the competition between local antiferromagnetic correlations, involving superexchange-coupled Cu moments, and the doped holes that would like to lower their kinetic energy by delocalizing.  The relative strength of the AF correlations decreases with doping, and the AF correlations become overdamped for $p>p_c$.  

There is also a small pseudogap that results from the dependence of the pairing scale on $p$ and the local variation in $p$ due to disorder and the long-range-Coulomb interaction.  This energy scale appears to be limited by the doping-dependent spin fluctuation spectrum, and in particular the energy $E_{\rm cross}$, corresponding to the neck of the hourglass dispersion.  The onset of bulk superconductivity is limited by regions with low-energy AF fluctuations; the magnitude of the spin gap established in the superconducting state corresponds to the energy scale for spatially-uniform superconductivity.  

\acknowledgments

I have benefited from discussions with and comments by K. Fujita, C. C. Homes, M. H\"ucker, P. D. Johnson, S. A. Kivelson,  P. A. Lee, Y. Li, H. Miao, N. J. Robinson, Senthil, and many others.  This work was supported at Brookhaven by the Office of Basic Energy Sciences, U.S. Department of Energy (DOE) under Contract No.\ DE-SC0012704.

\bibliography{LNO,theory}%,neutrons,misc}

\end{document}